\documentclass[aps,prl,twocolumn,showpacs,10pt]{revtex4-1}
\usepackage{graphicx,amsmath,amssymb,amsfonts,sidecap,color}

\newcommand{\e}[1]{{\rm e}^{#1}}

\newcommand{\ket}[1]{\ensuremath{\left|#1\right\rangle}}

\makeatletter
\newcommand*{\rom}[1]{\expandafter\@slowromancap\romannumeral #1@}
\makeatother

\begin{document}

\title{Parafermions in Interacting Nanowire Bundle}

\author{Jelena Klinovaja$^{1,2}$}
\author{Daniel Loss$^2$}
\affiliation{$^1$Department of Physics, Harvard University,  Cambridge, Massachusetts 02138, USA,}
\affiliation{$^2$Department of Physics, University of Basel,
             Klingelbergstrasse 82, CH-4056 Basel, Switzerland}
             
\date{\today}
\pacs{71.10.Pm; 05.30.Pr; 73.21.Hb; 71.10.Fd}
	
\begin{abstract}
We propose a scheme to induce $\mathbb{Z}_3$ parafermion modes, exotic zero-energy bound states that possess non-Abelian statistics. We consider a minimal setup consisting of a bundle of four tunnel coupled nanowires hosting spinless electrons that interact strongly with each other. The hallmark of our setup is that it relies only on simple one-dimensional wires, uniform magnetic fields, and strong interactions, but does not require the presence of superconductivity or exotic quantum Hall phases. 
\end{abstract}

\maketitle

{\it Introduction.} Topological properties in condensed matter systems have attracted  wide attention over the past decade \cite{Nayak,Alicea_2012}. In recent years, the focus has shifted to  ever more exotic localized states appearing at interfaces. Such quasiparticles are often protected from local perturbations, and thus are promising  candidates for qubits.
 In particular, the physics of Majorana fermions (MFs)
 has been explored in detail due to their promise of non-Abelian statistics \cite{Wilczek,Alicea_2012,Halperin_MF} useful for topological quantum computing \cite{Kitaev_2003,Nayak,Alicea_2012}. Several systems have been proposed to host MFs, such as fractional quantum Hall effect (FQHE) systems  \cite{Read_2000,Nayak}, topological insulators \cite{fu, Nagaosa_2009,Ando}, optical lattices \cite{Sato,demler_2011},  $p$-wave superconductors \cite{potter_majoranas_2011}, nanowires with strong Rashba spin orbit interaction \cite{lutchyn_majorana_wire_2010, oreg_majorana_wire_2010, alicea_majoranas_2010,mourik_signatures_2012,das_evidence_2012,deng_observation_2012, marcus_MF,Rokhinson,Goldhaber,Rotating_field}, self-tuning RKKY systems \cite{RKKY_Basel,RKKY_Simon,RKKY_Franz}, and  graphene-like systems \cite{Klinovaja_CNT, bilayer_MF_2012, MF_nanoribbon, MF_MOS_Zigzag, MF_MOS, MF_Bena}. Though MFs possess non-Abelian statistics, it is of  Ising type and thus not sufficient for  universal quantum computation
~\cite{Alicea_braiding, Halperin_braiding}.

The search for an ideal system to unambiguously observe MFs still continues. At the same time, a lot of effort is invested in identifying systems that host even more exotic quasiparticles that obey non-Abelian statistics of the Fibonacci type \cite{Cheng,PF_Linder,Vaezi,Ady_FMF,PF_Clarke,PF_Mong,twist_barkeshli,bilayer_PFs,vaezi_2}.
Generating such quasiparticles is a crucial  step towards a more  powerful braid statistics that  enables 
universal topological quantum computing. In contrast to MFs, proposals for Fibonacci anyons require the presence of strong electron-electron interactions. As a result, systems in the FQHE regime (which requires by itself ultraclean samples) seem to be attractive candidates: they host fractional excitations that serve as building blocks for Fibonacci anyons. However, another important ingredient for many setups is superconductivity, for which a strong magnetic field needed for FQHE has a detrimental effect. As a result, to make progress, one needs to search for schemes that do not require both strong magnetic fields and superconductivity \cite{twist_barkeshli,bilayer_PFs}. 

\begin{figure}[bt!]
\includegraphics[width=0.80\linewidth]{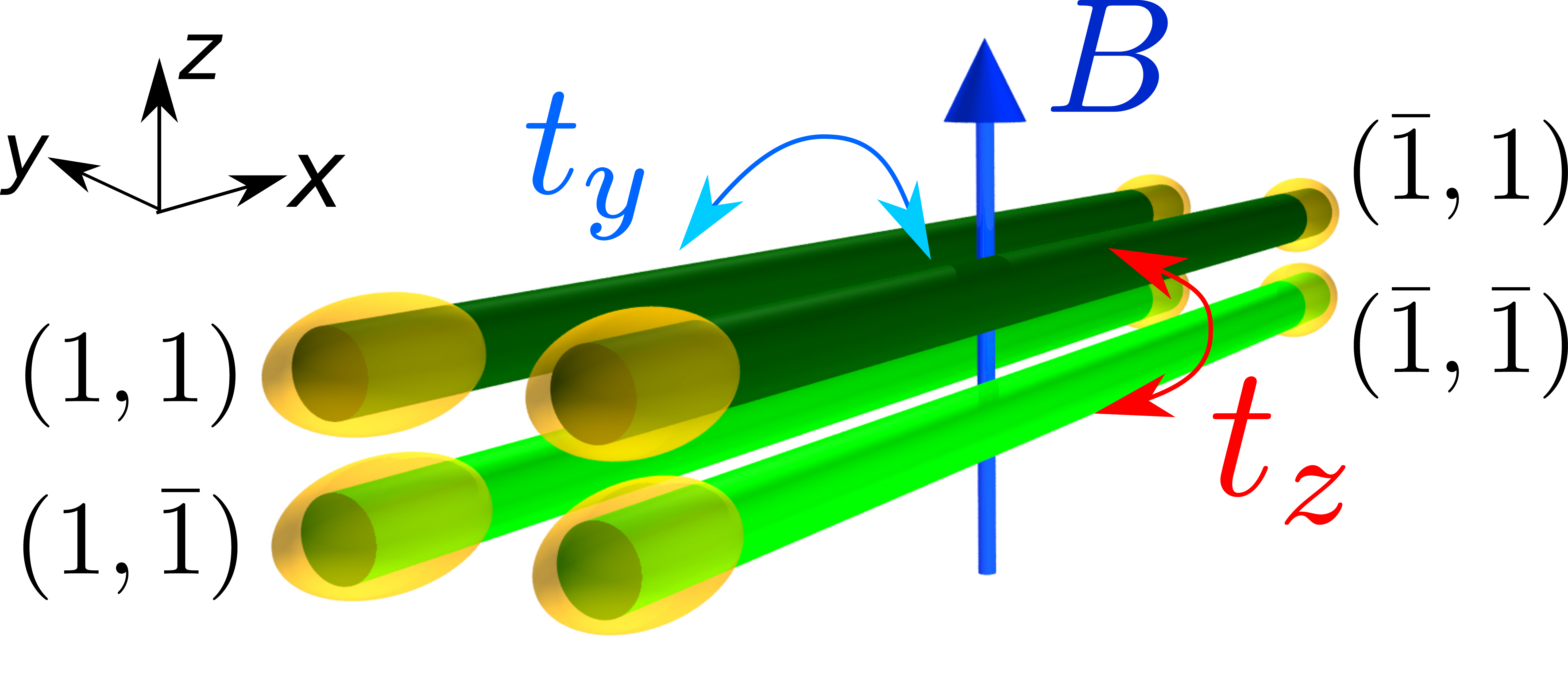}
\caption{Bundle of four one-dimensional nanowires (green cylinders) hosting spinless electrons and aligned in $x$ direction. The upper (dark green) and lower (light green) wires are tunnel coupled to each other by the hopping amplitude $t_z$. Similarly, the two upper (lower) wires are tunnel coupled by  $t_y$. A uniform magnetic field ${\bf B} = B \hat z$ (blue arrow) penetrating the bundle in $z$ direction results in   
the opening of Peierls gap.
 For strong electron-electron interactions,
 zero-energy PFs emerge at the ends (yellow spheres).  }
\label{bundle}
\end{figure}

In this Letter we propose a setup that neither relies  on superconductivity nor on  FQHE. 
Instead, we consider a bundle of four tunnel-coupled nanowires hosting spinless interacting electrons in the presence of a uniform magnetic field, see Fig. \ref{bundle}. The electrons at the upper (lower) wires are assumed to have positive (negative) masses. As a potential realization for such systems, we envisage, for example, spin polarized graphene nanoribbons \cite{MF_nanoribbon} and semiconducting nanowires, or optical lattices and cold atom systems \cite{Review_cold_atoms}. The advantage of one-dimensional wires lies in the easier control of boundaries between topological and non-topological phases \cite{Ady_FMF,Alicea_braiding}.

First, we demonstrate that even in the absence of  interactions, the system can be gapped and be in two topologically distinct phases. In the trivial phase, there are no bound states in the gap. In the topological phase, there are fractionally charged fermions of  Jackie-Rebbi type \cite
{Jackie_Rebbi,SSH,FF_suhas,Klinovaja_Loss_Ladder,FF_MF,FF_Diego} localized at the bundle ends. Such fermions are interesting on their own due to their non-Abelian statistics of Ising type \cite{Klinovaja_Loss_Ladder}. Second, if electron-electron interactions inside each wire are strong, we show that the system hosts so-called parafermions (PFs) \cite{Fradkin_PF_1980,Fendley_PF_2012}, sometimes also referred to as fractional MFs \cite{PF_Mong}. In particular, we focus on the case of $\mathbb{Z}_3$ PFs. For this case, we show threefold degeneracy of the ground state and construct the corresponding zero-mode operators satisfying $\mathbb{Z}_3$ PF statistics. Despite the fact that  such PFs still do not provide universal braid statistics, it is richer than that of MFs \cite{PF_Clarke,PF_Mong}.

{\it System.} We consider a bundle of four parallel wires of length $l$  aligned in the $x$ direction, see Fig. \ref{bundle}. The upper (lower) two wires form the upper (lower) ladder, which lies in the $xy$ plane. A uniform magnetic field ${\bf B} = B \hat z$ is applied along the $z$ direction, and the corresponding vector potential ${\bf A }$ is chosen to point along the $y$ direction, ${\bf A }=  (B x) \hat y$. Each wire is labeled by two indices $\tau$ and $\sigma$, where $\tau=1$ ($\tau=-1$) refers to the left (right) wires, and $\sigma=1$ ($\sigma=-1$) refers to the upper (lower) ladders. In addition, the two upper (lower) wires, possess a positive (negative) mass $m$ ($-m$). The chemical potentials $\mu_\sigma$ in each ladder are tuned in such a way that the Fermi wavevector $k_F$ is identical in all four wires, $\mu_1=-\mu_{\bar 1}\equiv \mu$. As a result, the kinetic part of the Hamiltonian becomes
\begin{equation}
H_0=\sum_{\tau,\sigma=\pm1}\int dx\ \sigma \Psi_{\tau\sigma}^\dagger(x)\left[-\frac{\hbar^2\partial_x^2}{2m}-\mu\right] \Psi_{\tau\sigma}(x),
\end{equation}
where $\Psi_{\tau\sigma}^\dagger(x)$ [$\Psi_{\tau\sigma}(x)$] is the electron creation (annihilation) operator at the position $x$ of the $(\tau,\sigma)$-wire.
 
The upper and lower wires are weakly tunnel coupled to each other,
described by the tunneling Hamiltonian
\begin{equation}
H_z=\sum_{\tau=\pm1}\int dx\ t_z \left[  \Psi_{\tau1}^\dagger(x)  \Psi_{\tau\bar{1}}(x) + H.c.\right],
\label{tz}
\end{equation}
with $t_z$ denoting the hopping amplitude.
The hopping in $y$-direction is accompanied by a magnetic Peierls phase $\varphi_B (x)=(e/\hbar c)\int d{\bf r}\ {\bf A \cdot r}\equiv (e B a_y/\hbar c)x$ \cite{footnote_1}, resulting in
\begin{equation}
H_y=\sum_{\sigma=\pm1}\int dx\ t_y \left[  \Psi_{1,\sigma}^\dagger(x)  e^{i \varphi_B (x) } \Psi_{\bar{1},\sigma}(x) + H.c.\right],
\end{equation}
where $t_y$ is the hopping amplitude between the upper or  lower two wires separated by a distance $a_y$. Without loss of generality, we assume that $t_y$ and $t_z$ are real and
positive.
Furthermore,  $t_y,t_z$ are independent of the position $x$, so that the momentum along the wire is conserved in the tunneling process (for $B=0$).

\begin{figure}[bt!]
\includegraphics[width=0.80\linewidth]{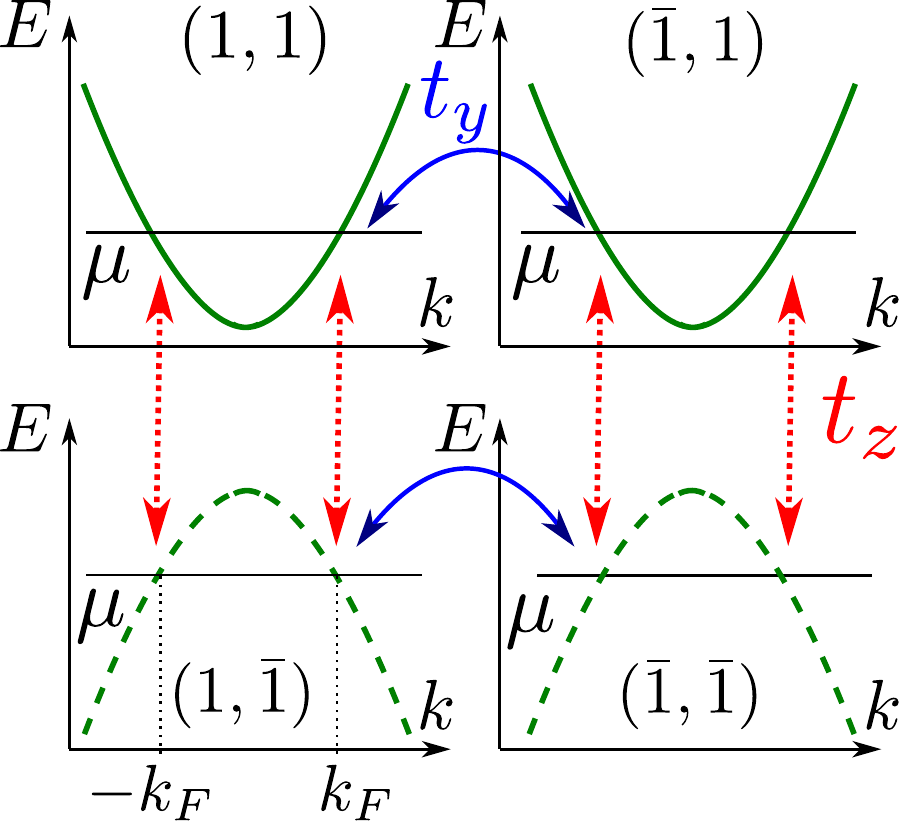}
\caption{The energy spectrum of the four wires. Each wire is labeled by two indices  ($\tau,\sigma$). The two upper (lower) wires with $\sigma=1$ ($\sigma=\bar 1$) have positive (negative) mass. The hopping of strength $t_z$ between upper and lower wires (red dashed arrows) is momentum-conserving. In contrast, the hopping of strength $t_y$ between left and right wires (blue solid arrows) results in scattering between two different momenta in the presence of a magnetic field $B$. In particular, at $B=B_1$ the momentum transfer is equal to $2k_F$, where $k_F$ is the Fermi momentum defined by the chemical potential $\mu$. Such resonant scattering between right and left mover fields leads to the opening of a Peierls gap at the Fermi level.  In addition, if $t_y>t_z$, there are zero-energy bound states localized at the ends of the system.}
\label{spectrum}
\end{figure}

{\it Direct resonant scattering.}
To find the energy spectrum of the system of four wires in the presence of the magnetic field, we linearize the total Hamiltonian $H=H_0+H_y+H_z$ around the Fermi points $\pm k_F$ by representing the electron operators in terms of slowly varying right ($R_{\tau\sigma}$) and left ($L_{\tau\sigma}$) movers \cite{Klinovaja2012, Klinovaja_Loss_Ladder},
\begin{align}
\Psi_{\tau\sigma}(x)=R_{\tau\sigma}(x)e^{i\sigma k_F x}+L_{\tau\sigma}(x) e^{-i\sigma k_Fx}.
\end{align}
First, we focus on the case of {\it direct} resonant scattering between the  upper and  lower two wires \cite{kane,QHE_Klinovaja_PRB,QHE_Klinovaja_PRL}. Such a resonance occurs when the magnetic phase $\varphi_B (x)$ provides a momentum transfer of $2k_F$ necessary for back scattering between $L_{\tau\sigma}$ and $R_{\tau\sigma}$ (so that the total momentum is conserved). This condition is satisfied at the magnetic field $B_1= 2k_F \hbar c /e a_y$ corresponding to the magnetic phase $\varphi_{B_1} (x)=2k_Fx$. We note that similar resonances are at the  origin of Peierls phase transition \cite{Braunecker,Peierls}.

In the new basis ($R_{11}$,$L_{11}$,$R_{ 1\bar 1}$,$L_{1\bar  1}$,$R_{\bar 11}$,$L_{\bar 11}$,$R_{\bar 1\bar 1}$,$L_{\bar 1\bar 1}$), the total Hamiltonian becomes
\begin{align}
H=\hbar\upsilon_F \hat k \eta_3 +\frac{t_y}{2}(\tau_1\eta_1-\sigma_3 \tau_2\eta_2)+t_z \sigma_1 \eta_1,
\end{align}
where the Pauli matrices $\tau_i$ ($\sigma_i$) act in left/right (upper/lower) wire space,
and the Pauli matrices $\eta_i$ act in right/left mover space. Here, $\hbar \hat k = -i \partial_x$ is the momentum operator with eigenvalues k taken from the corresponding Fermi points $\pm k_F$,
and $\upsilon_F = \hbar k_F/m$ is the Fermi velocity.
The energy spectrum of $H$ is given by
\begin{align}
&E^{(1)}_{\pm}=\pm\sqrt{(\hbar \upsilon_F k)^2+t_z^2},\\
&E^{(2)}_{\pm, \pm}=\pm\sqrt{(\hbar \upsilon_F k)^2+(t_z\pm t_y)^2},
\end{align}
where $E^{(1)}_{\pm}$ is twofold degenerate. We note that the system is gapless if  $t_y=\pm t_z$ and is fully gapped otherwise. This signals the presence of a topological phase transition that separates a topological phase with bound states inside the gap from a trivial phase without bound states. Indeed, we find that zero-energy bound states, one localized at the left end and one at the right end of the system, exist only if $t_y>t_z$. The wavefunction of the left bound state is given by
\begin{equation}
\Psi_{F}=\begin{pmatrix}
e^{-ik_Fx}\\i e^{-ik_Fx}\\-i e^{ik_Fx}\\-e^{ik_Fx}
\end{pmatrix} e^{-x/\xi_1}+\begin{pmatrix}-e^{ik_Fx}\\-ie^{ik_Fx}\\ie^{-ik_Fx}\\e^{-ik_Fx}
\end{pmatrix} e^{-x/\xi_2}
\label{bound_state}
\end{equation}
in the basis ($\Psi_{11}$, $\Psi_{1\bar 1}$, $\Psi_{\bar 11}$, $\Psi_{\bar 1\bar 1}$). Similarly, one can construct the wavefunction of the right bound state.  The localization length of the bound states  is inverse proportional to the smallest gap, $\xi={\rm max} \{\xi_1 , \xi_2 \}$, where  $\xi_1 = \hbar \upsilon_F/t_z$ and $\xi_2 = \hbar \upsilon_F/(t_y-t_z)$.

{\it Resonant scattering assisted by electron-electron interactions.} From now on, we focus on magnetic fields such that the momentum transfer due to the magnetic phase $\varphi_B$ is larger than $2k_F$, so that the resonant scattering between left and right movers is possible only in the presence of strong electron-electron interactions, where the excess moment is transferred back to the electron system with the assistance of back-scattering \cite{QHE_Klinovaja_PRL,QHE_Klinovaja_PRB,kane,kane_PRL,Ady_FMF}. The resonance occurs at  magnetic fields satisfying $B_n=n B_1$, when the momentum transfer $2 n k_F$ is commensurable with $2k_F$.  In what follows, we consider,  without loss of  generality, the magnetic field increased by a factor of three, i.e. $B_3=3B_1$. 

To account for electron-electron interactions, we take into the consideration the back scattering term of the strength  $g_B\propto U_{2k_F}$, where $U_{2k_F}$ is the interaction potential \cite{Giamarchi}. This allows us to construct the following momentum-conserving terms
\begin{align}
&{\mathcal O}_{t_y\sigma} =g_y(R_{\sigma\sigma}^\dagger L_{\bar{\sigma}\sigma})(R_{1\sigma}^\dagger L_{1\sigma})(R_{\bar 1\sigma}^\dagger L_{\bar{1}\sigma})+H.c.,
\label{interaction_y}
\end{align}
which describe scattering processes between the  upper ($\sigma=1$) or lower  ($\sigma=-1$) two wires. The momentum transfer provided by the magnetic field $B_3$ is $6k_F$. These $6k_F$ are distributed between the process of one electron tunneling between wires ($R_{\sigma\sigma}^\dagger L_{\bar{\sigma}\sigma}$) and the process of two electrons scattering off each other inside a left  ($R_{1\sigma}^\dagger L_{1\sigma}$) and a right ($R_{\bar 1\sigma}^\dagger L_{\bar{1}\sigma}$) wire. As a result, three left moving electrons are back-scattered into three right moving electrons. Here, the parameter $g_y$ is proportional to both the initial hopping matrix element $t_y$ and the back-scattering strength $g_B$, $g_y\propto t_y g_B^2$.
Next, we introduce chiral bosonic fields $\phi_{r\tau\sigma}$ such that
 $R_{\tau\sigma}=e^{i \phi_{1\tau\sigma}}$ and $L_{\tau\sigma}=e^{i \phi_{\bar 1\tau\sigma}}$. 
The chiral fields are subjected to the commutation relations
\begin{align}
\label{chiral1}
[\phi_{r\tau\sigma}(x), \phi_{r\tau\sigma}(x')]=i  \pi r  \ {\rm sgn}(x-x'),
\end{align}
with all other commutators vanishing.
With this choice  $R_{\tau\sigma}$ ($L_{\tau\sigma}$) satisfies the proper fermionic anticommutation relations,
while the remaining ones can be satisfied by an appropriate choice of Klein factors~\cite{Giamarchi}.
In terms of chiral fields, Eq.~(\ref{interaction_y})  becomes
${\mathcal O}_{t_y\sigma} =g_y \cos (2\phi_{1\sigma\sigma}+\phi_{1\bar\sigma\sigma}-2\phi_{\bar1\bar \sigma\sigma}-\phi_{\bar1 \sigma \sigma})$.
The tunneling in the $y$ direction induces coupling between all four fields belonging to the upper ($\sigma=1$) or lower  ($\sigma=\bar 1$) wires.  Similarly, the operators describing tunneling in the $z$ direction between upper and lower wires $O_{t_z\tau 1}=R_{\tau1}^\dagger L_{\tau\bar1}+H.c.$ and $O_{t_z\tau\bar1}=L_{\tau 1}^\dagger R_{\tau\bar1}+H.c.$ [see Eq. (\ref{tz})] become ${\mathcal O}_{t_z\tau r} = t_z\cos (\phi_{\bar r\tau\bar 1}-\phi_{r\tau 1})$.

We note that the sets  $\{{\mathcal O}_{t_y\sigma}\}$ and $\{{\mathcal O}_{t_z\tau r}\}$ do not commute with each other and, thus, cannot be diagonalized simultaneously, so only one of them can be relevant.
Their relative strengths can be assessed from their scaling dimensions~\cite{kane,Ady_FMF,Giamarchi}, which can be readily found 
in the representation of conjugate variables, $\phi_{r\tau\sigma}= [r \varphi_{\alpha}-\theta_{\alpha}+\tau (r \varphi_{\beta}-\theta_{\beta})+\sigma(r \varphi_{\gamma}
-\theta_{\gamma})+\tau \sigma (r \varphi_{\delta}-\theta_{\delta})]/2$. Thus, ${\mathcal O}_{t_y\sigma}$ scales with $K_y=[9(K_\alpha+K_\gamma)+K_\beta^{-1}+K_\delta^{-1}]/4$, and ${\mathcal O}_{t_z\tau r}$ scales with $K_{zz}=[K_\alpha+K_\beta+K_\gamma^{-1}+K_\delta^{-1}]/4$, where $K_j$ 
is the interaction parameter corresponding to 
$\varphi_j$ and $\theta_j$. 

In what follows, we work in the regime where ${\mathcal O}_{t_y\sigma}$ are relevant operators \cite{Ady_FMF}. This is the case if $K_y<K_{zz}$ or if the bare $t_z$ is of order one, so that it does not flow, and $K_y<2$. 
This would imply that in this regime the system is only partially gapped 
(by ${\mathcal O}_{t_y\sigma}$). However, there are additional terms due to interactions that we need to consider,
\begin{align}
&{\mathcal O}_{t_z\tau}=g_z (L_{\tau\tau}^\dagger  R_{\tau\bar{\tau}}) (L_{\tau1}^\dagger R_{\tau 1})(L_{\tau\bar 1}^\dagger  R_{\tau\bar{1}})+H.c.
\end{align}
Rewritten in terms of chiral fields, this becomes
${\mathcal O}_{t_z\tau}=g_z\cos (2\phi_{\bar 1\tau\tau}+\phi_{\bar 1\tau\bar \tau}-2\phi_{1\tau\bar \tau}-\phi_{1\tau\tau})$,
 with scaling dimensions $K_z=[9(K_\alpha+K_\beta)+(K_\gamma^{-1}+K_\delta^{-1})]/4$.
These operators correspond to momentum-conserving tunneling processes between upper and lower wires that also include back scattering  inside the wires, giving
again $g_z \propto t_z g_B^2$. 
Most importantly, the set  $\{{\mathcal O}_{t_y\sigma}\}$  commutes with $\{{\mathcal O}_{t_z\tau}\}$. Assuming that both ${\mathcal O}_{t_y\sigma}$ and ${\mathcal O}_{t_z\tau}$ are relevant operators, we work in the limit of large $g_y$ and $g_z$. As a result, the four commuting cosine operators fully gap the spectrum consisting of eight branches.
Consequently, there are no propagating states inside the gap but there could be bound states localized at the system ends similar to the non-interacting case, see Eq. (\ref{bound_state}).
This is indeed the case, as we show next.

{\it Parafermion bound states.} 
To simplify further analysis, we change the basis to $\eta_{r\tau \sigma}=2\phi_{r\tau\sigma}-\phi_{\bar r\tau\sigma}$, where
the four commuting terms become
\begin{align}
&{\mathcal O}_{t_y\sigma} =g_y \cos (\eta_{1 \sigma \sigma}-\eta_{\bar 1\bar \sigma \sigma}),\\
&{\mathcal O}_{t_z\tau}=g_z \cos(\eta_{\bar1\tau\tau}-\eta_{1\tau\bar \tau}).
\end{align}
Thus, we arrive at the conclusion that all fields are pinned pairwise:
$\eta_{1\sigma\sigma}(x)=\eta_{\bar 1\bar \sigma \sigma}(x)$ and $ \eta_{\bar1\tau\tau}(x)=\eta_{1\tau\bar \tau}(x)$,
so the system is fully gapped. 
 We note that the left and right movers are not independent due to the vanishing boundary conditions at  $x=0,\ell$, 
 {\it i.e.}, $R_{\tau\sigma}(x)=-L_{\tau\sigma}(-x)$, with $R_{\tau\sigma}(x)=R_{\tau\sigma}(x+2 \ell)$~\cite{LL_Eggert,LL_Gogolin}.
Consequently,
$\eta_{1\tau\sigma}(x=0,\ell)=\eta_{\bar 1\tau\sigma}(x=0,\ell)+\pi$.

Next, we  apply the unfolding procedure \cite{LL_Eggert,LL_Gogolin,Giamarchi,PF_Clarke,Ady_FMF,FF_suhas}. To halve the number of fields, we double the wire lengths, so that they extend now 
from  $-\ell$ to $\ell$. Moreover, new 
chiral fields  $\xi_{r\tau}(x)$ are defined in such a way that the boundary conditions are automatically satisfied,
\begin{align}
&\xi_{r\tau}(x)=\begin{cases}
    \eta_{r (r\tau) \tau}(x),&  x\in[0, \ell], \\
    \eta_{\bar r (r \tau) \tau}  (-x)  +\pi,        & x\in[ -\ell, 0].
\end{cases}
\label{chiralnew}
\end{align}
Here, the index $r=1$ ($r=\bar 1$) corresponds to the right (left) mover field, and the index $\tau=\pm 1$ distinguishes between two fields of the same chirality.
The commutation relations for $\xi_{r\tau}$
follow then from Eqs. (\ref{chiral1}) and (\ref{chiralnew}) with the Klein factors taken into account (see Supplemental Material~\cite{Supplemental}),
with the only non-vanishing commutator given by
\begin{equation}
[\xi_{r\tau}(x), \xi_{r\tau}(x')]=3ir\pi\  {\rm{ sgn}} (x-x').
 \label{com_1}
\end{equation} 
As a result, the operators ${\mathcal O}_{t_y\sigma}(x) =g_y(x) \cos(\xi_{R\tau}-\xi_{L\tau})$ are defined on the interval $[0,\ell]$ and  ${\mathcal O}_{t_z\tau} (x)=g_z(x) \cos(\xi_{R\tau}-\xi_{L\bar \tau})$ on  $[-\ell,0]$. 
 
In a final step, we introduce canonically conjugate fields $\phi_{1,2}$ and $\theta_{1,2}$, via
$\xi_{r\tau}= (r\phi_1 -  \theta_1 + 3 \tau r \phi_2 - 3 \tau \theta_2)/2$~\cite{Supplemental}.
The sum of operators with $g_y(x)$ having support in $[0,\ell]$ becomes
\begin{align}
&{\mathcal O}_{t_y1}+{\mathcal O}_{t_y\bar1}=
2 g_y(x) \cos (\phi_1) \cos (3\phi_2),
\end{align}
and, similarly, the ones with $g_z(x)$-support in $ [-\ell,0]$,
\begin{align}
&{\mathcal O}_{t_z1}+{\mathcal O}_{t_z\bar1}=
2 g_z(x) \cos (\phi_1) \cos (3\theta_2).
\end{align}
In the limit of large $g_y$ and $g_z$, the fields are pinned in both intervals to minimize the total energy,
\begin{align}
&\phi_1(x)=\pi {M} , \ \ x\in[-\ell, \ell], \\
&\phi_2(x)=\pi ({ M}+1+2{ n})/3, \ \ x\in[0, \ell], \\
&\theta_2(x)=\pi ({ M}+1+2{m})/3, \ \ x\in[ -\ell, 0],
\end{align}
where $M,m,n$ are integer valued operators.
First, we note that the field $\phi_1$ is pinned uniformly to minimize the kinetic energy.
Second, the non-commuting fields $\phi_2$ and $\theta_2$ are pinned in two neighboring regions separated by an (infinitesimal) interface. Thus, we come to a standard situation~\cite{Cheng,PF_Linder,Vaezi,Ady_FMF,PF_Clarke,PF_Mong,twist_barkeshli,bilayer_PFs,vaezi_2}
 where bound states arise at such interfaces.
Indeed, following Refs.~\cite{PF_Clarke,PF_Mong} we can construct two operators $\alpha_{1}$ and $\alpha_{\bar 1}$,
\begin{align}
\alpha_{1}=e^{i2\pi(m-n)/3},\ \alpha_{\bar 1}=e^{i2\pi(m+n)/3},
\end{align}
which commute with the Hamiltonian, and, hence, represent zero energy bound states, localized at $x=0$ and $x=\ell$.
First, we note the following properties,
\begin{align}
&\alpha_{1}^3=1,\  \alpha_{1}^2= \alpha_{1}^\dagger, \ \alpha_{\bar 1}^3=1,\ \alpha_{\bar 1}^2 = \alpha_{\bar 1}^\dagger.
\end{align}
Second, we use the non-trivial commutation relation $[\phi_{2}(x), \theta_{2}(x')]=-(i\pi/3){\rm sgn}(x-x')$,
which follows directly from 
Eq. (\ref{com_1}), giving
$[m,n]=3i/4\pi$.
From this, we finally arrive at the following  commutation relations between two zero-energy operators 
\begin{align}
&\alpha_{1} \alpha_{\bar 1} 
=e^{-2\pi i/3}\alpha_{\bar 1} \alpha_{1},\ \ \alpha_{1}^\dagger \alpha_{\bar 1} 
=e^{2\pi i/3}\alpha_{\bar 1} \alpha_{1}^\dagger.
\label{U_commutators}
\end{align}
This is our central result, which shows that $\alpha_{r}$ 
represent $\mathbb{Z}_3$  parafermion modes \cite{Fradkin_PF_1980,Fendley_PF_2012}.

Finally, we show that the ground state is threefold degenerate.
We note that $(\alpha^\dagger_{1} \alpha_{\bar 1})^3=1$, so $\alpha^\dagger_{1} \alpha_{\bar 1}$ has three distinct eigenvalues $e^{2i\pi q /3}$, where  
$q=0,\pm 1 \ ({\rm mod}\ 3) $.
The corresponding eigenstates are denoted by $\ket{q}$.
Using Eq. (\ref{U_commutators}) with an appropriate phase choice \cite{PF_Clarke},
we find
\begin{align}
\alpha_{1} \ket{q} = \ket{q+1}, \ \alpha_{1}^\dagger \ket{q} = \ket{q-1}, 
\end{align}
so the ground state  is threefold degenerate.

{\it Conclusions.} We proposed a  simple system that hosts zero-energy $\mathbb{Z}_3$ parafermions. The main motivation for our work was to construct a minimal setup that does not require the presence of superconductivity and/or exotic quantum Hall edge states. Instead, we consider a bundle of tunnel coupled nanowires in the presence of a magnetic field, and rely only on strong
interactions \cite{footnote_1}. 
To advance further to even more exotic modes, in particular  Fibonacci anyons, we envisage construction of two-dimensional networks consisting of wire bundles (similarly to Ref. \onlinecite{PF_Mong}). 
Evidently, there are many experimental challenges such as growing clean tunnel-coupled wires, controlling chemical potentials, {\it etc.}, but one might hope that the experimental progress for PFs will be as rapid as for MFs. In addition, the present Letter serves as a proof of principle that addresses an important question whether the quantum Hall effect and superconductivity are necessary ingredients. As we have shown, there are much larger classes of engineered setups at which one can aim.

\acknowledgments
We acknowledge valuable discussions with J. Alicea and A. Stern. DL thanks the UCSB KITP for hospitality.
This research is supported by the  Harvard Quantum Optics Center, the Swiss NSF, the NCCR QSIT, and the US NSF under Grant No. NSF PHY11-25915.

\appendix

\section{Appendix: Commutation relations}

In this Appendix we provide details on the derivation of the commutation relations for the chiral fields $\xi_{r\tau}$ defined in terms of $\eta_{r\tau\sigma}$-fields [see Eq. (\ref{chiralnew}) of the main text]. Because the unfolding procedure involves only fields belonging to the {\it same} $(\tau\sigma)$-wire, the commutation relations between fields belonging to {\it different} wires,
\begin{align}
&[\xi_{r\tau}(x), \xi_{\bar r\tau'}(x')]=[\xi_{r\tau}(x), \xi_{ r', \bar \tau }(x')]=0, \label{com_3}
\end{align}
are obtained straightforwardly from Eq. (\ref{chiral1}). Similarly, the commutation relation $[\xi_{r\tau}(x), \xi_{ r\tau}(x')]$, provided that $x$ and $x'$ are either both positive or both negative, are not affected by the unfolding procedure [see Eq. (\ref{chiralnew})].  
However, we should emphasize that if $x>0$ and $x'<0$, or vice versa, we need to come back to the fermionic operator language and cannot make direct use of Eqs. (\ref{chiral1}) and (\ref{chiralnew}) anymore, since otherwise we would arrive at the incorrect conclusion that the previous commutator would also vanish in this case (which it should not). The origin of this apparent
inconsistency  lies in the missing Klein factors responsible for giving the correct anticommutation relations between right and left mover fields. The unfolding procedure  performed in the fermionic language [$R_{\tau\sigma}(x)=-L_{\tau\sigma}(-x)$], where the bosonic operator $e^{i\eta_{1\tau\sigma} (x)}$ [$e^{i\eta_{\bar 1\tau\sigma}}(x')$] corresponds to the fermionic operator $R^\dagger_{\tau\sigma} (x)R^\dagger_{\tau\sigma} (x)L_{\tau\sigma} (x)$ [$L^\dagger_{\tau\sigma} (x')L^\dagger_{\tau\sigma}(x')R_{\tau\sigma}(x')$],
leads then to non-vanishing commutation relations for $\xi_{r\tau}(x)$ also for the case $xx'<0$.
Moreover, the commutation relations should be smooth around $x=0$. This additional requirement results in Eq. (\ref{com_1}),
\begin{align}
&[\xi_{r\tau}(x), \xi_{r\tau}(x')]=3ir\pi\  {\rm{ sgn}} (x-x').
\end{align}
To conclude, we also present the only non-zero commutation relations for $\phi_{1,2}$ and $\theta_{1,2}$ fields, defined via $\xi_{r\tau}=(r \phi_1 -\theta_1+3\tau r \phi_2 - 3\tau \theta_2)/2$,
\begin{align}
&[\phi_{1}(x), \theta_{1}(x')]= -3i\pi  {\rm{ sgn}} (x-x'),\\
&[\phi_{2}(x), \theta_{2}(x')]= -\frac{i\pi}{3}  {\rm{ sgn}} (x-x').
\end{align}


\begin{thebibliography}{100}
\bibitem{footnote_1}  We emphasize that our setup being based on the orbital effect from the magnetic field cannot be considered as a strictly one-dimensional system.
\end{thebibliography}
\end{document}